\documentstyle[multicol,aps,epsf]{revtex}
\begin{document}
\draft
\title{Interface roughening with nonlinear surface tension}
\author{Barbara Drossel}
\address{Physics Department, Massachusetts Institute of Technology, Cambridge, MA 02139, USA}
\date{\today}
\maketitle
\begin{abstract}
Using stability arguments, this Brief Report suggests that a term that enhances the surface tension in the presence of large height fluctuations should be included in the Kardar-Parisi-Zhang equation.  A  one-loop renormalization group analysis then shows for interface dimensions larger than $\simeq 3.3$ an unstable strong-coupling fixed point that enters the system from infinity. The relevance of these results to the roughening transition is discussed.
\end{abstract}

\begin{multicols}{2}
During recent years, kinetic roughening of growing interfaces has been 
the object of intense research. In 1986, Kardar, Parisi, and Zhang (KPZ) \cite{KPZ}
suggested the following Langevin equation to describe the macroscopic dynamics 
of a stochastically growing interface:
\begin{equation}
{\partial h\over \partial t} = \nu \nabla^2 h + {\lambda \over 2} (\nabla h)^2 + \eta\, . \label{KPZ}
\end{equation}
Here, $h = h(\vec x,t)$ is the (coarse-grained) height of the $d$-dimensional surface, and $\eta$ is a stochastic noise that roughens the interface. It is uncorrelated and Gaussian distributed, i.e. $\langle \eta(\vec x, t)\rangle = 0$ and
\begin{equation}
\langle \eta(\vec x, t) \eta(\vec x', t')\rangle = 2 D \delta^d(\vec x - \vec x') \delta(t - t')\, . 
\end{equation}
The first term on the right-hand side of Eq.~(\ref{KPZ}) is a surface tension that tends to smooth the interface; the second term accounts for growth perpendicular to the surface orientation. In principle, a constant term has to be added, which, however, can be absorbed in the definition of the height variable. 

A comprehensive discussion of Eq.~(\ref{KPZ}) can be found in \cite{KS,HHZ}.
Near the stationary state, the height profile is self-affine, i.e. invariant under a rescaling $h'(\vec x,t) = b^{-\chi} h(b\vec x, b^z t)$. The exponent $\chi$ is the roughness exponent, $z$ is the dynamical critical exponent. 
Eq.~(\ref{KPZ}) is invariant under the transformation $h'(\vec x,t) = h(\vec x + \lambda \vec v t,t) + \vec v \cdot \vec x$, that tilts the interface by a small angle $\vec v$ \cite{FNS}, leading to the exponent relation
\begin{equation}
\chi + z = 2\, . \label{scalingrelation}
\end{equation}

In one dimension, the critical exponents $\chi$ and $z$ are exactly known, since the stationary solution of the Fokker-Planck equation
\begin{eqnarray*}
{\partial P \over \partial t} &=& -\int_0^L dx {\delta\over \delta h}
\left\{\left[\nu {d^2h\over dx^2} + {\lambda \over 2} \left(dh\over dx\right)^2 \right] P \right\} \\
&&+ D\int_0^L dx {\delta^2 \over \delta h^2}P
\end{eqnarray*}
is given by
\begin{equation}
P[h(x)] = \exp\left[-{\nu\over 2D}\int_0^Ldx \left(dh\over dx\right)^2\right]\, ,
\label{stationary}
\end{equation}
leading to $\chi = 1/2$ and (with Eq.~(\ref{scalingrelation})) $z = 3/2$ (see e.g. \cite{K}. 

In higher dimensions, the stationary distribution is not exactly known, and the analytical approaches that have been taken so far, are renormalization group theory \cite{KPZ,MHKZ,TF} and self-consistent methods \cite{DMKB,BC,T,MBDBC,HF,FHT}. 
A renormalization group calculation to one-loop order gives the following flow equations for the parameters \cite{KPZ,MHKZ}
\begin{eqnarray}
{d\nu(l) \over dl} &=& \nu(l)\left(z - 2 + {2 - d \over d}g \right)\, , \nonumber \\
{dD(l) \over dl} &=& D(l) \left(z - d - 2\chi + g \right) \, , \label{flow}\\
{d\lambda(l) \over dl} &=&  \lambda(l)(\chi + z - 2) \,, \nonumber
\end{eqnarray}
where $g = K_d \Lambda^{d-2} \lambda^2 D / 4\nu^3$ is an effective coupling constant, with $K_d$ being the surface area of the $d$-dimensional unit sphere, divided by $(2\pi)^d$, and $\Lambda$ the cutoff for the wave vector. Equations (\ref{flow}) can be combined to form an equation for $g$ alone,
\begin{equation}
{dg\over dl} = \left(2 - d + {4d - 6 \over d} g\right)\, . \label{flowg}
\end{equation}
In one dimension, this equation has a stable fixed point at $g^* = 1/2$. Inserting the fixed-point value in the flow equations for $\nu$ and $D$ gives again the above-mentioned values for the critical exponents. The second fixed point $g^* = 0$ is unstable.

In 2 dimensions, the fixed point $g^* = 0$ is still unstable, and there is no stable fixed point at all. In dimensions larger than 2, the fixed point $g^*=0$ becomes stable, and an unstable fixed point $g^* = d(d-2)/(4d-6)$ appears that separates the region where the flow goes to zero from the region where it diverges. These results indicate that there exists a phase transition from a flat interface to  a rough interface for dimensions larger than 2. Although the rough phase is presumably controlled by an attractive fixed point, this fixed point is not accessible within a one-loop renormalization group. 
A two-loop calculation yields similar results \cite{TF}, and recently it has been shown that a finite stable fixed point is not reached by perturbative calculations to any order \cite{L}. 

There are certainly  many possible explanations for this failure of perturbation theory (e.g. lack of a small parameter, or nonanalytical behavior). In this Brief Report, I suggest a further explanation by noting that Eqs.~(\ref{flow}) might reflect a physical instability of Eq.~(\ref{KPZ}), and that a stabilizing term is necessary. Integrating the flow equation for $\nu$ (Eq.~(\ref{flow})) in the neighborhood 
of $\nu = 0$, one finds that $\nu$ flows through zero and becomes negative for some finite value of $l$ (see also \cite{Gol}). If this behavior reflects the physics of the system, it indicates that the interface is unstable, since the surface tension becomes negative. 
A self-consistent mean-field theory \cite{orland} for the KPZ equation is also unstable, suggesting an instability at least in high dimensions. A stable mean-field theory is presented in \cite{MB}, where a nonlinear surface tension is added to Eq.~(\ref{KPZ}). This mean-field theory leads to an interface that has bumps of a characteristic height. Instabilities have even  been found in computer simulations of the  (discretized) KPZ equation in two dimensions \cite{MB}. A continuum limit, however, is only well defined if the discrete model leads to a stable and smooth interface at large scales, and it seems therefore necessary to include a stabilizing term in  Eq.~(\ref{KPZ}).

The equation studied in this paper is the following \cite{MB}:
\begin{equation}
{\partial h\over \partial t} = \nu \nabla^2 h + {\lambda \over 2} (\nabla h)^2 + {\kappa\over 2} \nabla^2 h(\nabla h)^2 + \eta\, . \label{KPZ2}
\end{equation}
The new term proportional to $\kappa$ is generated at the second order in the expansion for the surface tension (see e.g. \cite{KS}) and is the stabilizing term in the above-mentioned mean-field theory \cite{MB}.
It is possible that the nonlinear surface tension, although present at microscopic scales, becomes irrelevant at larger scales, which is indeed the case in the neighborhood of the fixed points $g^* = 0$ and $g^* = d(d-2)/(4d-6)$, as we shall see below. We will, however, find an additional fixed point where $\kappa$ is not small, suggesting that the stabilizing term is important even at large scales in the rough phase. 

Let us first discuss the effect of the new term on the one-dimensional KPZ equation. The stationary height distribution is now
\begin{eqnarray}
P[h(x)] &=& \exp\Biggl\{-\int_0^Ldx \nonumber \\
&&\times\Biggl[{\nu\over 2D}\left(dh\over dx\right)^2
+{\kappa\over 24D} \left(dh\over dx\right)^4\Biggr]\Biggr\}\, .
\label{stationary2}
\end{eqnarray}
Under a rescaling $x\to x'=x/b$ and $h\to h'=h/b^\chi$ with $\chi = 1/2$, the second term is multiplied by $1/b$, while the first term remains invariant, indicating the irrelevance of $\kappa /D$. 
A one-loop renormalization group calculation for the stationary distribution in Eq.~(\ref{stationary2}) leads to the same conclusion. 
In momentum space, the propagator is $D/\nu k^2$, and the vertex is $-(\kappa/24 D) k_1k_2k_3k_4 \delta(k_1+k_2+k_3+k_4)$. To one-loop order, one obtains after a short calculation the following flow equations:
\begin{eqnarray}
{d(\nu/D)\over dl} &=& {\nu\over D}\left(2\chi-1+{\kappa D\over 2\nu^2} K_1\Lambda\right)\, ,\nonumber\\
{d(\kappa/D)\over dl} &=& {\kappa\over D}\left(4\chi-3-{3\kappa D\over 2\nu^2} K_1\Lambda\right)\, , \label{flow2static}
\end{eqnarray}
leading to
\begin{equation}
{dc\over dl} = c\left(-1-5c\right) \label{flowc}
\end{equation}
for the effective coupling constant $c=(\kappa D /2\nu^2) K_1\Lambda$. The fixed point $c^*=0$ is stable, indicating that the stabilizing term can be neglected at large length scales.
There is also an unstable fixed point at $c^* = -1/5$. For initial values $c<-1/5$, the flow of $c$ goes to $-\infty$, indicating that the surface becomes unstable in this parameter region. We will not consider further the region of negative $c$. Since the negative fixed point must also occur in the flow equations found from  dynamics, Eq.~(\ref{KPZ2}), it can however be used to make sure that the calculations do not contain errors. 

Let us now renormalize the equation of motion Eq.~(\ref{KPZ2}) to one-loop order. Inserting the equation of motion in the Gaussian probability distribution of the noise
\begin{equation}
W[\eta] \propto \exp\left[ -{1\over 4D} \int d^d x \int dt \, \eta(\vec x,t)^2\right]\, , \label{noise}
\end{equation}
and introducing an auxiliary field $\tilde h$, we obtain the weight of a given space-time configuration $[h(\vec x,t)]$ \cite{BJW}
\begin{displaymath}
W[h] \propto \int {\cal D} [i\tilde h] \exp\left\{{\cal J} [\tilde h,h]\right\}\,,
\end{displaymath}
with the dynamical functional
\begin{eqnarray}
{\cal J}[\tilde h, h] &= &\int d^d x \int dt \Biggl\{D \tilde h \tilde h \nonumber -\tilde h \times\\
&&\times \left[{\partial h\over \partial t} - \nu \nabla^2 h - {\lambda \over 2} (\nabla h)^2 - {\kappa \over 2} \nabla^2 h(\nabla h)^2  \right] \Biggr\}\,.
\label{functional}
\end{eqnarray}
The dynamical functional ${\cal J}$ plays the same role in dynamical renormalization group as the Hamiltonian in statics. The propagators of this model are
\begin{equation}
G_0(\vec k, t) \equiv \langle \tilde h(-\vec k, t) h(\vec k,t) \rangle_0 = \theta(t) e^{-\nu k^2 t} \label{resp}
\end{equation}
and
\begin{equation}
C_0(\vec k, t) \equiv \langle h(-\vec k, t) h(\vec k,t) \rangle_0 = D e^{-\nu k^2 |t|} /\nu k^2 \, ,\label{corr}
\end{equation}
and the vertices are
\begin{displaymath}
-{\lambda \over 2} (\vec k_2 \cdot \vec k_3) \tilde h_{\vec k_1} h_{\vec k_2} h_{\vec k_3} \delta(\vec k_1 + \vec k_2 + \vec k_3)
\end{displaymath}
and
\begin{displaymath}
{\kappa \over 2} k_2^2 (\vec k_3 \cdot \vec k_4) \tilde h_{\vec k_1} h_{\vec k_2} h_{\vec k_3} h_{\vec k_4} \delta(\vec k_1 + \vec k_2 + \vec k_3+ \vec k_4)\, .
\end{displaymath}
Here, $\langle \dots \rangle_0$ indicates an average within the Gaussian theory, where the nonlinear terms proportional to $\lambda$ and $\kappa$ are not taken into account. 
Renormalization of this model is done by first integrating over the large wave vectors $\Lambda/b < k < \Lambda$, where $\Lambda$ is the wave vector cutoff, and $b = 1 + l$ is close to 1. Next, the system is rescaled to the original size by introducing new variables $k' = bk$, $t' = t/b^z$, $h'=h/b^\chi$. To one-loop order, this gives the following flow equations for the parameters:
\begin{eqnarray}
{d\nu \over dl} &=& \nu\left(z - 2 + {\lambda^2D \over 4\nu^3}{2-d\over d} K_d\Lambda^{d-2} + {\kappa D \over 2\nu^2} K_d \Lambda^d\right)\,,\nonumber\\
{dD\over dl} &=& D\left(z-d-2\chi + K_d {\lambda^2D \over 4\nu^3} \Lambda^{d-2}\right) \,,\nonumber\\
{d\lambda \over dl} &=& \lambda\left(\chi + z - 2 -{\kappa D \over 2\nu^2}K_d \Lambda^d\right) \,,\nonumber \\
{d\kappa \over dl} &=& \kappa\Bigl(2\chi + z - 4 -{\kappa D \over 2\nu^2} {4-d\over d} K_d \Lambda^d \nonumber\\
&& \quad - {\lambda^2D \over 4\nu^3} {6-d\over d} K_d \Lambda^{d-2}\Bigr) \,,\label{flow2}
\end{eqnarray}
and finally for the effective coupling constants
$g=\lambda^2D K_d\Lambda^{d-2}/4\nu^3$
and
$c=\kappa D K_d \Lambda^d / 2\nu^2$,
\begin{eqnarray}
{dg\over dl} &=& g(2-d+{4d-6\over d}g-5c)\,, \nonumber\\
{dc\over dl} &=& c(-d+{4d-10\over d}g-{d+4\over d}c)\,.\label{flowgc}
\end{eqnarray}

\begin{figure}
\narrowtext
\centerline{\epsfysize=2.5in 
\epsffile{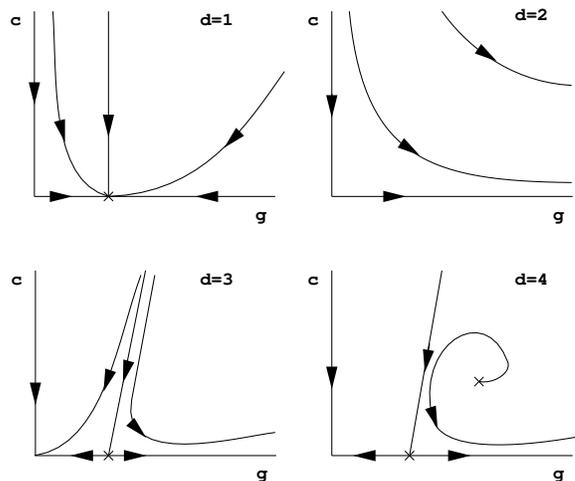}}
\caption{Schematic flow diagram obtained from Eq.~(16) in one, two, three, and four dimensions.}
\label{fig1}
\end{figure}

The diagrams contributing to $\kappa$ that contain four $\lambda$ vertices, cancel. This is a consequence of the tilt invariance of the KPZ equation: If $\kappa=0$, the system is tilt invariant, and this property must not change under rescaling. Therefore the flow diagram on the $g$-axis looks exactly as in the absence of $\kappa$. There is consequently a fixed point at $g^*=c^*=0$ that is unstable in one dimension, and stable above $d=2$. The fixed point $c^*=0$, $g^*=d(d-2)/(4d-6)$ is stable in $d<1.5$, and is a saddle point for $d>1.5$. A further fixed point is given by the intersection of the lines $2-d+g(4d-6)/d-5c=0$ and $-d+g(4d-10)/ d-c(d+4)/d =0$. For $d>3.295$, this fixed point lies in the physically interesting region $g>0$, $c>0$, and is given by $g^*=d(2d^2-d+4)/(8d^2-30d+12)$ and $c^*=(6d^2-10d)/(8d^2-30d+12)$. A linear stability analysis reveals that this fixed point is unstable, with two complex eigenvalues in dimensions between 3.574 and 11.53, and with two positive eigenvalues in all other dimensions. 
For dimensions close to 3.295, this fixed point is very large, indicating that it does not split from one of the other fixed points, but that it enters the system from infinity. The qualitative flow diagrams in  one, two, three, and four dimensions are shown in Fig.~\ref{fig1}. 

The values of the critical exponents $z$ and $\chi$ at the new fixed point are $z=(2d^3+5d^2-44d+16) / (8d^2-30d+12)$ and $\chi = (-2d^3+17d^2-26d+8)/(8d^2-30d+12)$. 
This one-loop result is certainly still far from reflecting the true behavior of the system, with a stable strong-coupling fixed point that occurs already in two dimensions, and with $0 \le \chi < 1$ and $z<d$. Nevertheless, we can draw some interesting conclusions: (i) In the presence of the stabilizing term, a fixed point can be found in the strong-coupling regime. Since this fixed point is fully repulsive in one-loop approximation, its presence does not affect the behavior of the system at large scales. However, it is possible that this fixed point becomes attractive and occurs already in two dimensions, when a better approximation is used. (ii) The strong-coupling fixed point enters the system from infinity and is not close to  the $g$ axis. If this is not just an artifact of the perturbation theory but reflects the true behavior of the system, there is no reason to expect that other higher-order terms can be neglected. (iii) The stabilizing term  and other terms that do not preserve the tilt invariance of the KPZ equation, cannot be generated by the  renormalization procedure. They have to be included explicitely in the beginning. (iv) The mapping of the KPZ equation on directed polymers in random media via the Cole-Hopf transformation \cite{K,KS,HHZ} cannot be performed any more in the presence of the nonlinear surface tension. Nevertheless, more direct mappings between growing Eden clusters and directed polymers \cite{rou91} suggest that the two systems should be equivalent even when higher-order terms are taken into account. Alternatively, as proposed in \cite{MB}, one might ask the question if the strong-coupling behavior can at all be characterized by a single universality class.

Calculations to higher loop orders or the inclusion of further terms might produce better results than the one-loop calculation presented here, but they are uncontrolled as long as the size of the neglected terms cannot be estimated. 
Self-consistent methods might be more successful than perturbation theory.
Mode-coupling theory has been shown to yield excellent results in one dimension, since the neglected vertex corrections are small \cite{HF,FHT}. It still has to be seen if this holds also in higher dimensions. 

I thank Mehran Kardar for countless discussions and for comments on the manuscript. I also thank E. Frey, H. Orland, and U. T\"auber for discussions.
This work was supported by the Deutsche 
Forschungsgemeinschaft (DFG) under Contract No. Dr 300/1-1.

\end{multicols}

\begin{references}{}
\bibitem{KPZ} M. Kardar, G. Parisi, and Y.-C. Zhang, Phys. Rev. Lett. {\bf 56}, 889 (1986).
\bibitem{KS} J. Krug and H. Spohn, in {\it Solids Far From Equilibrium: Growth, Morphology, and Defects,} ed. C. Godriche (Cambridge University Press, Cambridge, 1992).
\bibitem{HHZ} T. Halpin-Healy and Y.-C. Zhang, Phys. Rep. {\bf 254}, 215 (1995).\bibitem{FNS} D. Forster, D. Nelson, and M.J. Stephen, Phys. Rev. A {\bf 16}, 732 (1977). 
\bibitem{K} 
M. Kardar, Tr. J. of Physics, 18, 221 (1994).
\bibitem{MHKZ} E. Medina, T. Hwa, M. Kardar, and Y.-C. Zhang, Phys. Rev. A {\bf 39}, 3053 (1989). 
\bibitem{TF}  E. Frey and U.C. T\"auber, Phys. Rev. E {\bf 50} 1024 (1994).
\bibitem{DMKB} J.P. Doherty, M.A. Moore, J.M. Kim, and A.J. Bray, Phys. Rev. Lett. {\bf 72}, 2041 (1994).
\bibitem{BC} J.-P. Bouchaud, M. E. Cates, Phys. Rev. E {\bf 47}, 1455 (1993); {\it ibid.} {\bf 48}, 635 (1993). 
\bibitem{T} Y. Tu, Phys. Rev. Lett. {\bf 73}, 3109 (1994). 
\bibitem{MBDBC}  M.A.  Moore, T. Blum, J.P. Doherty, J.-P. Bouchaud, and P. Claudin, Phys. Rev. Lett. {\bf 74}, 4257 (1995).
\bibitem{HF} T. Hwa and E. Frey, Phys. Rev. A. {\bf 44}, R7873 (1991). 
\bibitem{FHT} E. Frey, U.C. T\"auber, and T. Hwa, Phys. Rev. E {\bf 53}, 4424 (1996). 
\bibitem{L} M. L\"assig, Nucl. Phys. B {\bf 448}, 559 (1995).
\bibitem{Gol} L. Golubovic and R. Bruinsma, Phys. Rev. Lett. {\bf 66}, 321 (1991). 
\bibitem{orland} H. Orland, private communication.
\bibitem{MB} M. Marsili and A.J. Bray, Phys. Rev. Lett. {\bf 76}, 2750 (1996).
\bibitem{BJW} R. Bausch, H. K. Janssen, H. Wagner, Z. Phys. B {\bf 24}, 113 (1976).
\bibitem{rou91}  S. Roux, A. Hansen, and E. L. Hinrichsen, 
                 J. Phys. A{\bf 24}, L295 (1991).
\end{references}
\end{document}